\documentclass[twocolumn]{revtex4-2}
\usepackage{graphicx}
\usepackage{dcolumn}
\usepackage{bm}
\usepackage{amsmath}
\usepackage{amsfonts}
\usepackage{amssymb}
\usepackage{color}
\usepackage{epstopdf}
\usepackage{float}
\providecommand{\U}[1]{\protect\rule{.1in}{.1in}}
\newcommand{\ket}[1]{|#1\rangle}
\newcommand{\bra}[1]{\langle#1|}

\begin{document}
\title{Evolution of expected values in open quantum systems.}
\author{Andr\'es Vallejo}
\author{Alejandro Romanelli}
\author{Virginia Feldman}
\author{Ra\'ul Donangelo}
\affiliation{\begin{small} Facultad de Ingenier\'{\i}a, 
Universidad de la Rep\'ublica, Montevideo, Uruguay\end{small}}
\date{\today}

\begin{abstract}
We derive a generalization of Ehrenfest theorem valid for open quantum systems. From this result, we identify three contributions to the evolution of expected values: i) the explicit time dependence of the observable, ii) the incompatibility between the observable and an operator which plays the role of an effective Hamiltonian, and iii) entropy changes. Considering the local Hamiltonian as the observable, and adopting a specific interpretation of the nature of thermal interactions, we obtain an alternative version of the first law of thermodynamics.
Within this framework, we show that in some cases the power performed by the system can be considered as a quantum observable. As an application, the pure dephasing process is reinterpreted from this perspective.


\end{abstract}

\maketitle

%
\section{\label{sec:level1}Introduction}

A well-known result in quantum theory states that, for a system 
in the state $\ket{\psi}$, the time derivative of the expected 
value of any observable quantity $O$ is given by the expression 
\cite{Messiah,Cohen}:
	\begin{equation}\label{dX_unitary}
	\dfrac{d\langle O\rangle}{dt}= \left\langle \dfrac{\partial O}{\partial t}\right\rangle +i\left\langle[H,O]			\right\rangle,
\end{equation}
where we have taken $\hbar=1$, $\langle O\rangle =\bra{\psi}O\ket{\psi}$
represents the expected value of the operator $O$, 
$H$ is the Hamiltonian of the system, and $[H,O]$ is the commutator 
between $H$ and $O$. 
This result has important applications. 
For instance, when applied to the position and momentum operators $X$ and $P$ 
of a particle with mass $m$, subjected to the potential $V(X)$, it leads to 
the famous \textit{Ehrenfest's Theorem} \cite{Ehrenfest}:
	\begin{equation}
	\dfrac{d\langle X\rangle}{dt}=\dfrac{\langle P\rangle}{m}, 
	\hspace{0.3cm}\dfrac{d\langle P\rangle}{dt}=-\left		
	\langle \dfrac{\partial V}{\partial X}\right\rangle ,
	\end{equation}
which, in some cases, provides support to the correspondence principle 
\cite{Hall}. 
Eq.~(\ref{dX_unitary}) is also the starting point for the derivation of 
important results involving the quantum speed limit \cite{Mandelstam}. 

Another obvious but relevant consequence of Eq.~(\ref{dX_unitary}) is 
that any time-independent observable that commutes with the Hamiltonian 
of a closed system is a constant of motion. 
In particular, the internal energy, defined as the expected value 
of the Hamiltonian,
	\begin{equation}
	E=\langle H\rangle
	\end{equation}
is a conserved quantity, provided that $H$ does not depend explicitly on time. 
This is consistent with the fact that closed classical systems have constant energy. 
However, when open systems are considered, the energy of
the system is no longer a conserved quantity due to the exchange of heat 
and work with the environment. 
In that case, it would be 
interesting to evaluate the effects of these interactions in the dynamics 
of expected values. 
In what follows, we derive a generalization of Eq.~(\ref{dX_unitary}) valid 
for an arbitrary open dynamics that allows us to address this issue.

The outline of this paper is as follows. 
In Section II we obtain the generalization of Eq.~(\ref{dX_unitary}) 
for the case of a generic open dynamics. 
In Section III, we show how the application of this result to the 
particular case in which the observable is the Hamiltonian of the system
leads, under some assumptions, to an alternative version of the first law of thermodynamics. 
We then focus on the study of 
two-level systems and show that the power of coherence can be interpreted as 
rotational work per unit of time in the presence of an external field. We also 
discuss the dephasing process from this perspective.
Remarks and conclusions are presented in Section IV.

\section{Time evolution of expected values in an arbitrary open dynamics}
The statistical properties of an open system are described by its reduced 
density operator $\rho$, in terms of which the expected value of any 
observable quantity $O$ can be written as:
	\begin{equation}\label{expected value}
	\langle O\rangle =\text{tr}\left(\rho O\right),
	\end{equation}
where \text{tr} denotes the trace operation. 
Let us consider the spectral decomposition of $\rho$,
	\begin{equation}\label{rho}
	\rho =\sum_{j}\lambda_{j}\ket{\psi_{j}}\bra{\psi_{j}},
	\end{equation}
where $\ket{\psi_{j}}$ is the eigenvector of $\rho$ corresponding to the 
eigenvalue $\lambda_{j}$.
From Eqs.~(\ref{expected value}) and (\ref{rho}), 
we obtain that
	\begin{equation}
	\langle O\rangle =\sum_{j}\lambda_{j}\bra{\psi_{j}}O\ket{\psi_{j}},
	\end{equation}
so an infinitesimal change in the expected value can be written as
	\begin{equation}\label{dX}
	d\langle O\rangle =\sum_{j}(d\lambda_{j})\bra{\psi_{j}}O\ket{\psi_{j}}+\sum_{j}\lambda_{j}
	d(\bra{\psi_{j}}O\ket{\psi_{j}}).
	\end{equation}

Given that $\rho$ is Hermitian, its set of eigenvectors evolves in time 
preserving the orthonormal-basis structure. 
Consequently, there exists a unitary operator $T$, possibly time-dependent, 
such that, for each eigenvector of $\rho$, satisfies that
	\begin{equation}\label{operatorT}
	\ket{\psi_{j}(t+dt)}=T(t,t+dt)\ket{\psi_{j}(t)}.
	\end{equation}
Since $d$t is small, $T$ is a unitary operator close to the identity, so 
we can approximate it by 
\cite{Lauvergnat}:
	\begin{equation}\label{operatorP}
	T(t,t+dt)=\mathbb{I}-i\Omega(t)dt,
	\end{equation}
where $\mathbb{I}$ is the identity operator and $\Omega$ is the Hermitian 
generator of the eigenbasis's dynamics. 
Using Eqs. (\ref{operatorT}) and (\ref{operatorP}), we obtain that
	\begin{equation}\label{dpsi}
		\begin{cases}
		d\ket{\psi_{j}}=\ket{\psi_{j}(t+dt)}-\ket{\psi_{j}(t)}=
		-i\Omega\ket{\psi_{j}}dt\\
		d\bra{\psi_{j}}=\bra{\psi_{j}(t+dt)}-\bra{\psi_{j}(t)}=
		i\bra{\psi_{j}}\Omega dt,
		\end{cases}
	\end{equation}
so from Eqs. (\ref{dX}) and (\ref{dpsi}), the variation in the expected 
value of $O$ can be written as
	\begin{equation}
		\begin{split}
		d\langle O\rangle 
		&=\sum_{j}d\lambda_{j}\bra{\psi_{j}}O\ket{\psi_{j}}+
		\sum_{j}\lambda_{j}\bra{\psi_{j}}dO\ket{\psi_{j}}\\
		&+i\sum_{j}\lambda_{j}\bra{\psi_{j}}[\Omega,O]\ket{\psi_{j}}dt.
		\end{split}
	\end{equation}
Therefore the time-derivative of $\langle O\rangle$ adopts the form
	\begin{equation}\label{dX2}
		\begin{split}
		\dfrac{d\langle O\rangle}{dt} 
		&= \sum_{j}\dfrac{d\lambda_{j}}{dt}\bra{\psi_{j}}O\ket{\psi_{j}}+
		\left\langle \dfrac{\partial O}{\partial t}\right\rangle 
		+i\left\langle[\Omega, O]\right\rangle,
		\end{split}
	\end{equation}
which is a natural generalization of Eq.~(\ref{dX_unitary}) for open systems.

When comparing Eqs. (\ref{dX_unitary}) and (\ref{dX2}), we note
two important differences. 
First, the appearance of the term
	\begin{equation}\label{dX_entropy}
	\sum_{j}\dfrac{d\lambda_{j}}{dt}\bra{\psi_{j}}O\ket{\psi_{j}},
	\end{equation}
which involves the time derivative of the eigenvalues of $\rho$. Since the evolution of 
a closed system is unitary, and noting that unitarity preserves the natural populations, we conclude that this contribution 
is necessarily related to some kind of interaction with the environment. One possible interpretation of this 
term arises recalling that the von Neumann entropy of the system can be written as
\begin{equation}
S_{vN}=-k_{B}\sum_{j}\lambda_{j}\ln\lambda_{j},
\end{equation}
where $k_B$ is the Boltzmann constant, so its rate of change adopts the form
	\begin{equation}\label{dS/dt}
	\dfrac{d{S}_{vN}}{dt}=-k_{B}\sum_{j}\dfrac{d\lambda_{j}}{dt}\ln\lambda_{j}.
	\end{equation}
Eqs. (\ref{dX_entropy}) and (\ref{dS/dt}) are clearly connected (albeit in an intricate way), 
since both are linear combinations of the set $\lbrace d\lambda_{j}/dt\rbrace$. Hence, under the hypothesis  
that the von Neumann entropy is a valid extension of the thermodynamic entropy, and considering 
that the contribution given by Eq. (\ref{dX_entropy}) is present only when there is an entropy variation, we conclude 
that it could be considered a contribution of thermodynamic nature, 
linked to the heat exchanged between the system and the environment.

On the other hand, the remaining terms in Eq.~(\ref{dX2}):
	\begin{equation}\label{dX_mec}
	\left\langle \frac{\partial O}{\partial t}\right\rangle 
	+i\left\langle[\Omega, O]\right\rangle 
	\end{equation}
correspond to those appearing in Eq.~(\ref{dX_unitary}), but with $\Omega$ 
playing the role of $H$. 
This is so because it is $\Omega$, instead of $H$, that governs the 
evolution of the eigenvectors of $\rho$ in an open dynamics. The first part
is clearly associated with the possible explicit time dependence of the observable, 
for example due to experimental control. On the other hand, it can be shown that a 
necessary condition for the second term to be non-zero is the presence of coherence 
in both the  $O$ and $\Omega$ eigenbases, as demonstrated later through an example.

\section{Energetic Implications}

Applying the previous formalism to the particular case in which the 
observable $O$ is the local Hamiltonian of the system, and defining the internal energy as the 
expected value of the Hamiltonian, $E=\langle H\rangle$, from Eq. (\ref{dX2}) we obtain that 
	\begin{equation}\label{dE1}
	\dfrac{dE}{dt} =\sum_{j}\dfrac{d\lambda_{j}}{dt}\bra{\psi_{j}}H\ket{\psi_{j}}+
	\left\langle \mathbb{P}_{H}\right\rangle,
	\end{equation}
where $\mathbb{P}_{H}$ is the Hermitian operator:
	\begin{equation}\label{power operator}
	\mathbb{P}_{H}=\dfrac{\partial H}{\partial t}+i[\Omega, H].
	\end{equation}

In light of the discussion carried out in the previous section, it 
is clear that the first term on the right-hand side of Eq.~(\ref{dE1}) 
corresponds to the rate of change of the internal energy that is 
accompanied by an entropy variation, so we interpret this term as the heat flux per unit of time exchanged by the system:
	\begin{equation}\label{heat}
	\dot{\mathcal{Q}}\equiv\sum_{j}\dfrac{d\lambda_{j}}{dt}
	\bra{\psi_{j}}H\ket{\psi_{j}},
	\end{equation} 
as it was recently proposed in Refs. \cite{Alipour1, Ahmadi, de Lima}.

Consequently, assuming that the classical statement of the first law
\begin{equation} \frac{dE}{dt}=\dot{\mathcal{Q}}+\dot{\mathcal{W}}
\end{equation}
still holds, we conclude that the second term in Eq.~(\ref{dE1}) is the total power performed on the system:
	\begin{equation}\label{work}
	\dot{\mathcal{W}}\equiv\langle\mathbb{P}_{H}\rangle.
	\end{equation}
The power includes the standard mechanical contribution $\langle\partial H/\partial t\rangle$ due to Hamiltonian driving, plus the \textit{coherence power} $i\langle [\Omega, H]\rangle$, which is related to 
the level of incompatibility between the local Hamiltonian $H$, which 
governs the free evolution of the system, and the effective Hamiltonian 
$\Omega$ that governs the evolution of the eigenbasis of $\rho$ when the 
interactions with the environment are included. Contrary to intuition, 
it is important to emphasize that $\Omega$ and $H$ are not directly related 
a priori. Instead, $\Omega$ depends on the global Hamiltonian, not solely 
on the local Hamiltonian $H$. Accordingly, the coherence power is 
the work per unit of time needed to deviate the quantum trajectory from the natural 
evolution defined by $H$ when the system is isolated from the environment.

Naturally, if the rotation of the eigenstates of $\rho$ is not modified 
by the interaction with the environment, $\Omega$ and $H$ commute. Therefore, 
no additional power is required to produce the new evolution. 
However, in the general case, both evolutions will be different, and since 
the local Hamiltonian $H$ fixes the natural evolution of 
the state of the system (and therefore, of the eigenbasis), deviations 
from that behaviour may imply an energy exchange
with the environment, which can be evaluated as the expected value of 
the commutator between both operators. 
 
In the case where both $H$ and $\Omega$ are time-independent, the total power corresponds 
to the expected value of the time-independent Hermitian operator $\mathbb{P}_{H}$, so 
it can be considered as a quantum observable. For this reason we will call  $\mathbb{P}_{H}$ 
\textit{the power operator}.

\section{Example: a two-level system}
\subsection{Physical interpretation of coherence work}

The density operator of a two-level system can be written in terms of the Bloch vector, $\vec{B}=\langle\vec{\sigma}\rangle$, 
as
	\begin{equation}\label{C1}
	\rho=\dfrac{1}{2}(\mathbb{I}+\vec{B}.\vec{\sigma}),
	\end{equation}
where $\vec{\sigma}$ is a formal vector whose components are the Pauli matrices, and 
$\mathbb{I}$ is the identity operator. Note that the components of $\vec{B}$ are 
proportional to the expected values of the spin operators, so it can be interpreted as a (dimensionless) 
microscopic magnetic dipole \cite{Nielsen}.

Similarly, the operators $\Omega$ and $H$ 
can be written, aside from irrelevant terms which are multiples of the identity, as:
	\begin{equation}\label{Omega}
	\Omega =-\vec{w}.\vec{\sigma},
	\end{equation}
and
	\begin{equation}\label{H}
	H =-\vec{v}.\vec{\sigma}.
	\end{equation}
In what follows, we assume that $H$ ($\vec{v}$) is time-independent. 

From Eqs.~(\ref{power operator}), (\ref{work}), (\ref{C1}), (\ref{Omega}), (\ref{H}), 
and employing the cyclic property of the triple product, it is possible to show that
	\begin{equation}\label{power2level1}
	\dot{\mathcal{W}}=i\left\langle[\Omega,H]\right\rangle=
	2\vec{w}.(\vec{B}\times\vec{v}).
	\end{equation}
To interpret Eq. (\ref{power2level1}) physically, notice that the Hamiltonian $H$ 
in Eq.~(\ref{H}) is equivalent to that describing the evolution of a spin in a magnetic field proportional 
to $\vec{v}$. As a consequence, the product  $\vec{B}\times\vec{v}$ represents the torque performed by the magnetic field $\vec{v}$ on the spin $\vec{B}$:
	\begin{equation}\label{torque}
	\vec{\tau}=\vec{B}\times\vec{v}.
	\end{equation}

Let us now focus on the other factor appearing in Eq. (\ref{power2level1}), $2\vec{w}$. Since $\vec{w}$ determines 
via Eq. (\ref{Omega}) the effective Hamiltonian $\Omega$ that governs the evolution of the eigenbasis of $\rho$ in Hilbert space, it is clear that $\vec{w}$ should be related to the evolution of the state in the Bloch vector representation. To clarify this point, let us note that the evolution of the normalized Bloch vector can be  understood as an instantaneous rotation around certain vector $\vec{\Omega}$, which plays the role of an angular velocity:

\begin{equation}\label{dB/dt}
	\frac{d\hat{B}}{dt}=\vec{\Omega}\times\hat{B}.
\end{equation}
From Ec. (\ref{H}) we can write the internal energy 
of the system as
\begin{equation}
E=\langle H\rangle =-\vec{B}.\vec{v},
\end{equation}
so the rate of change of the internal energy can be separated in two terms:
\begin{equation}\label{dE'}
\dfrac{dE}{dt}=-\dfrac{dB}{dt}\hat{B}.\vec{v} -B\dfrac{d\hat{B}}{dt}.\vec{v}
\end{equation}
Taking into account that the entropy of a two-level system depends only on the modulus of the Bloch vector, 
it was shown in Ref. \cite{Vallejo2021} that the first term in the right-hand side of the equation above is the heat exchanged with the environment per unit of time:
	\begin{equation}\label{heat2level}
	\dot{\mathcal{Q}}=-\dfrac{dB}{dt}(\hat{B}.\vec{v}),
	\end{equation}
Consequently, the second term represents the power performed by the system:
	\begin{equation}
	\dot{\mathcal{W}}=-B\dfrac{d\hat{B}}{dt}.\vec{v},
	\end{equation}
which, using Eq. (\ref{torque}) and (\ref{dB/dt}), can be written as:

	\begin{equation}\label{power2level2}
	\dot{\mathcal{W}}=-\vec{\Omega}.\vec{\tau},
	\end{equation}
so it adopts the classical form of the power needed to 
rotate a spin in the presence of an external field. Comparing Eqs. (\ref{power2level1}) and (\ref{power2level2}), we conclude that $-2\vec{w}$ can be identified as the angular velocity $\vec{\Omega}$. 

This simple model allows to clarify some aspects previously discussed 
in a more general way. 
First, notice that if the free angular velocity of the Bloch vector, which is proportional to $\vec{v}$, and the actual angular velocity, which is proportional to $\vec{w}$
are parallel, i.e. if $H$ and $\Omega$ commute,
the vectors $\vec{\tau}=\vec{B}\times\vec{v}$ and $\vec{\Omega}$ are 
orthogonal, so no external power is required to produce the evolution, 
see Fig.~(1). 
On the other hand, it is clear that the maximum power occurs when 
$\vec{\tau}$ and $\vec{\Omega}$ are antiparallel. 

As expected, this contribution only arises if the system exhibits coherence in the energy eigenbasis, as for incoherent states, the vectors $\vec{v}$ and $\vec{B}$ are parallel, which leads to zero torque according to Eq. (\ref{torque}). Furthermore, as shown in Eq. (\ref{power2level1}), the coherence power for a two-level system can be expressed as a triple product, which geometrically represents the volume of the parallelepiped defined by the vectors $2\vec{w}$, $\vec{B}$, and $\vec{v}$. Therefore, the coherence power is non-zero only if these vectors form a linearly independent set. Since two operators commute when their defining vectors are collinear, we conclude that, for the coherence power to be non-zero, none of the corresponding operators can commute with each other. This implies that, in particular, $\rho$ does not commute with either $H$ or $\Omega$, and thus must exhibit coherence in the eigenbases of both operators. These results are consistent with the findings in Ref. \cite{de Lima}, where the role of coherence in the first law is demonstrated and an equivalent expression for the energetic contribution of coherence is obtained.

\begin{figure}
\centering
{\includegraphics[width=1.0\columnwidth]{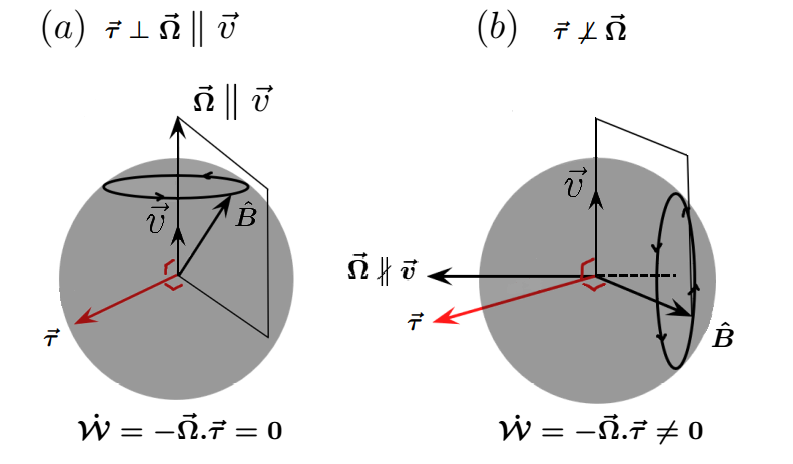}}
\caption{Geometrical interpretation of the coherence power for a 
two-level system. In the natural unitary evolution (left), the Bloch vector rotates around 
$\vec{\Omega}$, which is parallel to $\vec{v}$, so the vectors 
$\vec{\tau}$ and $\vec{\Omega}$ are orthogonal, and no power is required 
to maintain the rotation. 
If the vectors $\vec{\tau}$ and $\vec{\Omega}$ are not orthogonal (right), 
coherence power
is needed to deflect the trajectory from its natural evolution.}
\label{f1}
\end{figure}

\subsection{Alternative analysis of a pure-dephasing process}

As an example, let us analyze a pure-dephasing process. 
An outline of this discussion was presented in Ref. \cite{Vallejo2021}, 
and is now extended.

The pure-dephasing process, also called \textit{phase-damping}, is a 
quantum channel commonly employed as a schematic description
of the decoherence processes. 
In this process it is assumed that the coherence loss of an initial 
quantum superposition occurs due to interactions with the environment. 

This environment is supposed to be composed of small subsystems, e.g. 
photons from some background radiation, which are scattered by the system at a certain 
rate $\Gamma$. 
If we assume that the time scale during which the energy and momentum 
transfer is significant exceeds $\Gamma^{-1}$, it can be shown 
that the process is characterized by the preservation of the populations 
of the system, in parallel with an exponential decay of the coherences 
of its density matrix, when it is expressed in 
the energy eigenbasis \cite{Preskill}. 

Let us consider the pure-dephasing of a two-level 
atom with local Hamiltonian $H=-\varepsilon\sigma_{z}$,
corresponding to an effective field $\vec{v}=\varepsilon\hat{z}$. 
If the initial state is defined by the Bloch vector  $\Vec{B}(t=0)=(B_x,0,B_z)$, its evolution is given by \cite{Gong}:
	\begin{equation}\label{Bloch_PD}
	\vec{B}(t)=\left( B_x\cos(\Lambda t)
	e^{-\Gamma t},B_x\sin(\Lambda t)e^{-\Gamma t},B_z \right),
	\end{equation}
where $\Lambda=\sqrt{2\varepsilon}$.
From Eq.~(\ref{Bloch_PD}) we notice that the projection of the Bloch 
vector on the $\hat{z}$ direction is constant, so the energy of the 
atom does not change. 
Considering that the Hamiltonian of the system is constant, from the 
standard point of view there is no work involved in the process \cite{Alicki}. 
As a consequence, due to energy conservation, there is no heat exchanged either, so the information about the phase is lost without the 
intermediation of heat or work.  

\begin{figure}[h!]
\centering
{\includegraphics[width=0.85\columnwidth]{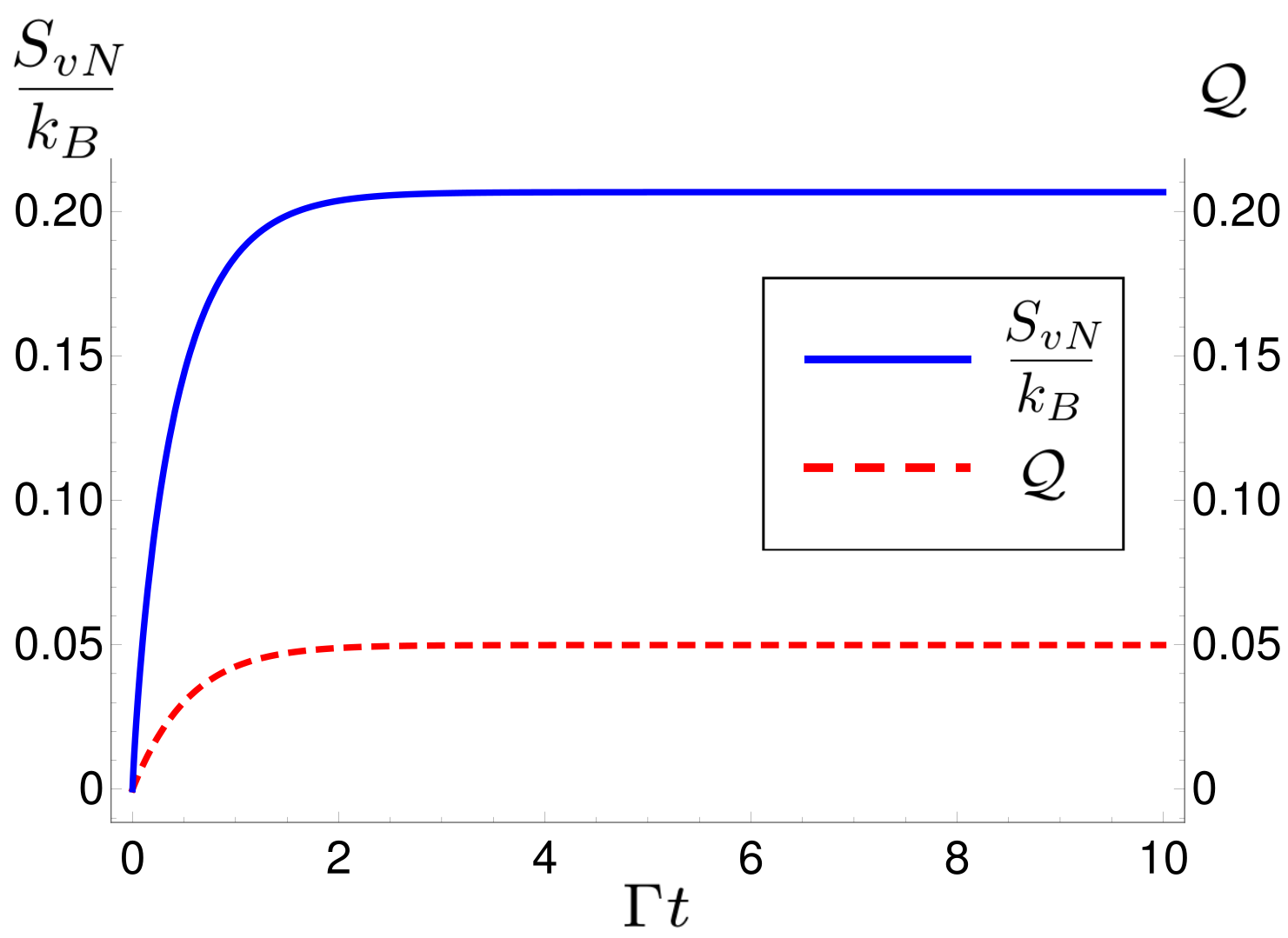}}
\caption{Evolution of the von Neumann entropy and the heat exchanged by an atom undergoing a pure dephasing process. In this framework, the entropy increase results from the absorption of heat from the environment. The parameter values are $B_x=\sqrt{0.2}$, $B_z=\sqrt{0.8}$, $\Lambda/\Gamma=0.4$, and $\varepsilon =0.5$}
\label{f2}
\end{figure}

However, the entropy increase occurring in the pure-dephasing process 
should imply that heat is absorbed 
from the environment, see Fig. (\ref{f2}). Integrating Eq.~(\ref{heat2level}) 
along the trajectory given by Eq.~(\ref{Bloch_PD}), we obtain:
	\begin{equation}
	\mathcal{Q}=\int_{0}^{\infty}\dot{\mathcal{Q}}dt=-\varepsilon 
	B_z\int_{1}^{B_z}\dfrac{dB}{B}=-\varepsilon B_z\ln\vert B_z\vert,
	\end{equation}
which, in fact, is a positive quantity for positive temperature states ($0<B_z<1$). Since the energy is constant, the system does an equivalent 
amount of coherence work on the environment, which should be associated to the scattering process. 

\vspace{0.2cm}

\section{Final Remarks and conclusions}

Since unitary evolutions preserve the eigenvalues 
of the density operator, the explicit time-dependence of these quantities 
is an accurate signature of the interaction of the system with its environment. 
In this work he have assumed that this interaction consists of a thermal 
contribution, which is associated with the entropy variation, and an 
additional contribution that does not change the entropy of the system. 
Noting that the entropy depends exclusively on those eigenvalues, we were able 
to determine to what extent these interactions affect the values of the other 
properties of the system, generalizing Ehrenfest's theorem, which is valid for 
unitary dynamics.

The above partition is based on the hypothesis that the von Neumann entropy is 
a proper extension of the thermodynamic entropy in the quantum regime. While 
this is strictly true for the particular case of Gibbs states, there is controversy 
about the validity of this identification for out-of-equilibrium states, in which 
the thermodynamic entropy is not even well-defined \cite{Hemmo,Prunkl,Chua}.
However, we believe that exploring the consequences of accepting that hypothesis 
is not only a valid conceptual exercise, but also a path that can contribute to 
shed light on this issue. In any case, note that the validity of Eq. (\ref{dX2}) 
is independent on this hypothesis, which only plays a role in the thermodynamic 
interpretation of the different terms appearing in that equation.

In addition to the thermal contribution and the possible time-dependence of observables 
through variable parameters, we have shown that the evolution of expected values is 
influenced by a coherence contribution linked to the level of incompatibility between 
the observables and a certain effective Hamiltonian $\Omega$ that governs the evolution of the 
eigenvectors of the density matrix. This operator, in two dimensions, is 
directly related to the angular velocity of the Bloch vector. Therefore, in the 
general case, it can be associated with the abstract angular velocity of the eigenbasis
of the density operator, which, due to the orthonormality condition, undergoes a 
rigid rotation in Hilbert space \cite{Synge}. 

As we already commented, $\Omega$ is 
not directly related to the bare Hamiltonian of the system. Furthermore, it might be 
tempting to assume that, for systems governed by a Lindblad equation under the Markovian 
approximation, $\Omega$ represents the effective Hamiltonian $H_{\text{eff}}$ appearing in the 
commutator term. This is generally not the case, as the evolution 
of the eigenstates of the density operator depends not only on $H_{\text{eff}}$ but also on the 
dissipative term of the equation. In this sense, the common claim that $H_{\text{eff}}$ appearing 
in the Lindblad equation is the generator of the unitary part of the evolution can be misleading, 
as it refers to aspects other than describing the unitary rotation of the eigenstates. It is 
possible to prove, however, that any reduced density operator satisfies a Lindblad-like equation 
in which $\Omega$, indeed, is the operator that appears within the commutator (generally,
this is not the Lindblad equation derived from the global Hamiltonian using standard methods). 
This observation has significant implications for designing shortcuts to adiabaticity, as discussed in Ref. 
\cite{Alipour2020}. 

For the particular case in which the observable considered is the 
Hamiltonian of the system, the application of the generalized Ehrenfest theorem 
gives rise to alternative versions of heat and work. 
However, Eq. (\ref{dX2}) can be interpreted as an infinite 
family of “first laws” applicable to different observables. In each case, the 
entropy-related term represents the thermal contribution to the change in the 
expected value of the corresponding observable. 
Under this interpretation, the first law of thermodynamics emerges as a specific 
instance within this broader framework.

Particularly interesting is the fact that for time-independent Hamiltonians, 
if the interaction is such that there is also no time dependence in the 
effective Hamiltonian $\Omega$, the power can be calculated as the expected 
value of a time independent Hermitian operator, so it can be considered 
as a quantum observable. 
Additionally, for two-level systems we have shown that the coherence power 
is related to the deviation of the Bloch vector motion with respect to its
natural tendency to rotate around the privileged direction determined 
by the external field.

We emphasize that several very recent works employ 
definitions of heat and work which are equivalent to Eqs. (\ref{heat2level}) 
and (\ref{power2level2}), but derived in the different context of shortcuts 
to adiabacity instead of as particular case of a generalized Ehrenfest formula 
\cite{Alipour1}. For example, in Ref. \cite{Choquehuanca}, entropy-based 
definitions of heat and work are employed to quantify the non-Markovianity of 
quantum dynamical maps, and a physical interpretation of coherence work 
(also called \textit{environment-induced work}) in terms of the ergotropy 
variation was reported in Ref. \cite{Choquehuanca2024}.

This interpretation, which is closer to Clausius' original spirit, allows us 
to analyze well-known processes, such as thermalization, or pure dephasing,
from an alternative point of view. 

\section*{Acknowledgments}

This work was partially supported by the Uruguayan agencies
Comisi\'on Acad\'emica de Posgrado (CAP), 
Agencia Nacional de Investigaci\'on e Innovaci\'on (ANII) and 
Programa de Desarrollo de las Ciencias B\'asicas (PEDECIBA).

\end{document}